\documentclass[runningheads]{cl2emult}

\usepackage{makeidx}  
\usepackage{graphicx} 
\usepackage{subeqnar} 
\usepackage{multicol} 
\usepackage{cropmark} 
\usepackage{eso}      
\makeindex            

\newcommand{\ignore}[1]{}
\newcommand{\beq}[1]{\begin{equation}\label{#1}}
\newcommand{\eeq}{\end{equation}}
\newcommand{\beqa}[1]{\begin{equation}\label{#1}\begin{eqalign}}
\newcommand{\eeqa}{\end{eqalign}\end{equation}}
\newcommand{\bsubeq}[1]{\begin{subequations}\label{#1}\begin{eqalignno}}
\newcommand{\esubeq}{\end{eqalignno}\end{subequations}}

\def \G   {{ G}}
\def \g   {{ g}} 

\def \C   {{ C}}
\def \c   {{ c}}

\def \Lm  {{ \Lambda}}
\def \l {{ \lambda}}

\def \L  {{\cal L}}
\def \bb  {{\beta^{-1}}}

\newcommand{\be}{\begin{equation}}
\newcommand{\ee}{\end{equation}}
\newcommand{\ba}{\begin{eqnarray}}
\newcommand{\ea}{\end{eqnarray}}

\newcommand{\nn}{\nonumber\\}

\newcommand{\bC}{{\bf C}}
\newcommand{\bF}{{\bf F}}

\newcommand{\bbf}{{\bf b}}

\newcommand{\bn}{{\bf n}}

\newcommand{\bx}{{\bf x}}

\newcommand{\bbmu}{\mbox{\boldmath $\mu$}}

\def\gs{\mathrel{\raise1.16pt\hbox{$>$}\kern-7.0pt 
\lower3.06pt\hbox{{$\scriptstyle \sim$}}}}         
\def\ls{\mathrel{\raise1.16pt\hbox{$<$}\kern-7.0pt 
\lower3.06pt\hbox{{$\scriptstyle \sim$}}}}         

\begin{document}
\title*
{Compression and Classification Methods 
\protect\newline 
for Galaxy  Spectra in Large Redshift Surveys}
\toctitle{Compression and Classification Methods 
\protect\newline 
for Galaxy Spectra in Large Redshift Surveys}

%
%
\titlerunning{Compression and Classification Methods}
%
\author{Ofer Lahav}
\authorrunning{Ofer Lahav}
%
%
\institute{Institute of Astronomy, Cambridge University,\\
Madingley Road, Cambridge CB3 0HA, UK}

\maketitle              

\begin{abstract}
Methods for compression and classification of galaxy spectra,
which are useful for large galaxy redshift surveys (such as the SDSS,
2dF, 6dF and VIRMOS), are reviewed. In particular, we describe and
contrast three methods: (i) Principal Component Analysis, (ii)
Information Bottleneck, and (iii) Fisher Matrix.  We show applications
to 2dF galaxy spectra and to mock semi-analytic spectra, and we
discuss how these methods can be used to study physical processes of
galaxy formation, clustering and galaxy biasing in the new large
redshift surveys.

\end{abstract}

\section{Introduction}

The classification of galaxies is commonly done 
using galaxy images, 
in the spirit of Hubble's original diagram and its extensions 
(for a review see van den Bergh 1998).
Galaxy spectra offer another way of classifying galaxies,
which can be directly connected to the underlying astrophysics. 
Obviously, the relation between galaxy morphology and spectra
also provides important insight into scenarios of galaxy formation.
One motivation for studying galaxy spectra in a statistical way is that 
new redshift surveys (e.g. SDSS, 2dF, 6dF and VIRMOS) 
will soon produce millions of spectra.
These large data sets can then be divided into subsets
for studies of e.g. luminosity functions and clustering per spectral type
(or a specific astrophysical parameter).
Traditional methods of
classifying galaxies ``by eye'' are clearly impractical in this
context. The analysis and full exploitation of such data sets require
well justified, automated and objective techniques to extract as much
information as possible.

The concept of spectral classification goes back to 
Humason (1936) and Morgan \& Mayall (1957).
The end goal of galaxy classification is a better understanding of the
physical origin of different populations and how they relate to one
another. In order to interpret the results of any objective
classification algorithm, we must relate the derived classes to the
physical and observable galaxy properties that are intuitively
familiar to astronomers. For example, important properties in
determining the spectral characteristics of a galaxy are its mean
stellar age and metallicity, or more generally its full star formation
history. 
An assumed star formation history can be translated into a synthetic
spectrum using models of stellar evolution (e.g., Bruzual \& Charlot
1993, 1996; Fioc \& Rocca-Volmerange 1997). 
Spectral features are also affected by
dust reddening and nebular emission lines.

As with other Astronomical data, there are different approaches for
analysing galaxy spectra. 
Conceptually, it is helpful to  distinguish between 
three procedures:  
\begin{itemize} 

\item Data compression

\item Classification 

\item Parameter estimation 

\end{itemize} 
The three might of course be related (e.g. classification can be done in a
compressed space of the spectra, 
or in the space of astrophysical parameters estimated from the spectra). 

Another distinction is between 'unsupervised' methods
(where the data `speak for
themselves', in a model-independent way) and 'supervised' methods
(based on training sets of models, or other data sets).
The statistical methods can also be viewed as a `bridge' between 
the data and the models, i.e. the same statistic can be applied
to both data and models, as an effective way of comparing the two.

The outline of this review is as follows: 
In Section 2
we mention briefly two examples of data sets (2dF spectra and mock spectra),
and then in Sections 3,4 and 5 we present three methods: 
PCA, the Information Bottleneck and Fisher Matrix.
In Section 6 
we compare and contrast these and some other methods,
and suggest directions for future work.

\section{Spectral Ensembles}
\label{sec:data}

We mention above the 
exponential growth of data of galaxy spectra.
Here we present 
two specific `proto-types' of real and mock data,
which are used later to illustrate the methods.

\subsection{Observed Spectra from the 2dF Survey}
\label{sec:data:2df}
The 2dF Galaxy Redshift Survey (2dFGRS; Colless 1998, Folkes et al. 1999) is
a major new redshift survey utilising the 2dF multi-fibre spectrograph
on the Anglo-Australian Telescope (AAT). The observational goal of the
survey is to obtain high quality spectra and redshifts for 250,000
galaxies to an extinction-corrected limit of $b_J$=19.45. The survey
will eventually cover approximately 2000 sq deg, made up of two
continuous declination strips plus 100 random $2^\circ$-diameter
fields.  Over 135,000 galaxy spectra have been obtained as of October 
2000.  The spectral scale is 4.3\AA~ per pixel and the FWHM resolution
is about 2 pixels.  Galaxies at the survey limit of $b_J$=19.45 have a
median S/N of $\sim 14$, which is more than adequate for measuring
redshifts and permits reliable spectral types to be determined.

Here we use a subset of 2dF galaxy spectra, previously used in the
analysis of Folkes et al. (1999).  The spectra are
given in terms of {\em photon counts} (as opposed to energy flux). The
spectra were de-redshifted to their rest frame and re-sampled to a
uniform spectral scale with 4\AA\ bins, from  3700\AA\ to 6650\AA. 
The  sample contains 5869 galaxies,
each described by 738 spectral bins. Throughout this paper, we refer
to this ensemble as the ``2dF sample''.
We corrected each spectrum by dividing it by a global system response
function (Folkes et al. 1999).  However, it is known that due to
various problems related to the telescope optics, the seeing, the
fibre aperture etc.  the above correction is not perfect. In fact,
each spectrum should be corrected by an individual response function
(work in progress).

  We note that another selection effect is due to the fixed diameter
of the 2dF fibre (of 2 arcsec, which corresponds in an Einstein-de
Sitter universe to $\sim 2.5 h^{-1}$ kpc at the survey median redshift
of 0.1).  The observed spectrum is hence sensitive to the fraction of
bulge versus disk which is in the fibre beam, and hence it it affected
by the distance to the galaxy (see Kochanek et al. 2000). 
However, other effects such as poor
seeing reduce this effect.  The `aperture bias' is also likely to be
less dramatic if the spectral diagnostic used is continuum-based
(rather than a diagnostic which is sensitive to emission lines that
originate from star-forming regions).  We also note that in a flux
limited sample, distant objects are more intrinsically luminous, and
this effect slightly biases distant populations towards early-type.

\subsection{Model Spectra from Semi-Analytic Hierarchical Merger Models}
\label{sec:data:models}

In one example of applying PCA to mock spectra
(Ronen, Aragon-Salamance \& Lahav 1999), the
star formation history was parameterized as a simple single burst or
an exponentially decreasing star formation rate. However, the
construction of the ensemble of galaxy spectra was done in an ad-hoc
manner.  An improvement to this approach is to use  a cosmologically
motivated ensemble of synthetic galaxies, with realistic star
formation histories. These histories are determined by the physical
processes of galaxy formation in the context of hierarchical structure
formation.  
Semi-analytic models have the advantage of
being computationally efficient, while being set within the
fashionable hierarchical framework of the Cold Dark Matter (CDM)
scenario of structure formation. In addition to model spectra, this
approach provides many physical properties of the galaxies, such as
the mean stellar age and metallicity, size, mass, bulge-to-disk ratio,
etc. This allows us to determine how effectively a given method can
extract this type of information from the spectra, which are
determined in a self-consistent way. In Slonim et al. (2000; herafter SSTL) 
we describe a ``mock 2dF'' sample, produced using the semi-analytic model
developed by Somerville (1997) and Somerville \& Primack (1999),
which has been shown to give good
agreement with many properties of local and high-redshift galaxies. 
The  ``mock 2dF'' given in SSTL has  $2611$ model galaxies with 
the same magnitude limit, wavelength coverage and spectral resolution, and
redshift range as the 2dF survey.
The effects of the response function of the fibres, aperture
effects, and systematic errors related to the placement of fibres 
were neglected.

The star formation histories were convolved with stellar population
models to calculate magnitudes and colors and produce model
spectra. SSTL  used the multi-metallicity GISSEL models (Bruzual \&
Charlot, in preparation) with a Salpeter IMF to calculate the stellar part of
the spectra. Emission lines from ionized $H_{\rm II}$ regions were
added using the empirical library included in the PEGASE models
(Fioc \& Rocca-Volmerange 1997). 
Dust extinction was included using an approach similar to that of
Guiderdoni \& Rocca-Volmerange (1987). 
Poisson noise was added, based on an empirical relation from the 2dF data.

\begin{figure}
\centering
\includegraphics[width=0.8\textwidth]{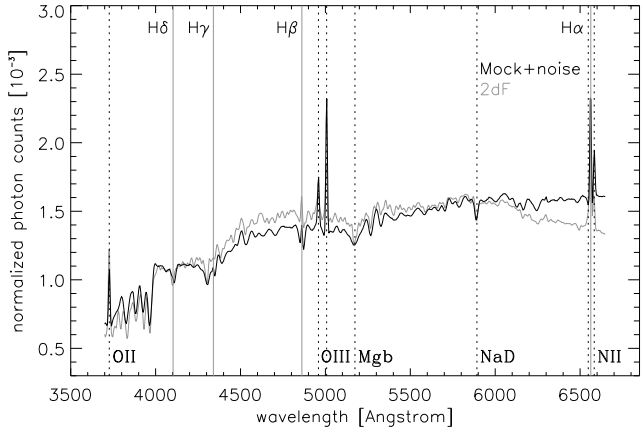}
\caption[]{Photon counts (sum normalised to unity) 
as a function of wavelength, averaged over the
entire catalogues of 2dF and mock+noise spectra. Note the familiar
spectral features such as the Balmer break at 4000 \AA, the Balmer
series (marked with vertical grey lines), and metal lines such as O,
Mg, Na, and N (marked by dotted lines).
One can see that the 2dF spectrum
appears to bend downwards relative to the models towards both ends of
the wavelength range. This may be due to an inaccurate correction for
the response function. 
(Figure from SSTL 2000.)
\label{fig:meanspec}}
\end{figure}

Figure~\ref{fig:meanspec} shows the mean spectrum for the 2dF and
mock+noise catalogues, obtained by simply averaging the photon counts
in each wavelength bin for all the galaxies in the ensemble. The mean
spectra for the observed and mock catalogues are seen to be
similar. The magnitude limit that we have chosen is such that our
ensembles are dominated by fairly bright, moderately star-forming
spiral galaxies, and the mean spectra show familiar features such as
the 4000 \AA\ break, the Balmer series, and metal lines such as
O$_{\rm II}$ and O$_{\rm III}$.

\section {Principal Component Analysis (PCA)}

PCA has previously been applied to data compression and classification
of spectral data of stars (e.g. Murtagh \& Heck 1987; Bailer-Jones et
al. 1997), QSO (e.g. Francis et al. 1992) and galaxies (e.g. Connolly
et al. 1995a; Folkes, Lahav \& Maddox 1996; Sodre \& Cuevas 1997; Galaz
\& de Lapparent 1997; Bromley et al. 1998; Glazebrook, Offer \&
Deeley 1998; Ronen et al. 1999; Folkes et al. 1999).  
While PCA operates as an efficient data compression
algorithm, it is purely linear, based only on the variance of the
distribution.  PCA on its own does not provide a rule for how to
divide the galaxies into classes
\footnote{
For example, in Folkes et al. (1999) the classification was done by
drawing lines in the $PC1-PC2$ plane using training sets.  One
training set was based on visual inspection of the spectra by a human
expert. This led to classification which is more sensitive to emission
and absorption lines, rather than to the continuum.}.

A spectrum, like any other vector, can be thought of as 
a point in a high-dimensional parameter space. One may
wish for a more compact description of the data.
By identifying the {\em linear}
combination of input parameters with maximum variance, PCA finds the
principal components that can be most effectively used
to characterise the inputs.

The formulation of standard PCA is as follows. Consider a set of $N_G$
galaxies ($i=1,N_G$), each with $N_S$ spectral bins ($j=1,N_S$). 
If $r_{ij}$ are
the original measurements of these parameters for these objects, then
mean subtracted quantities can be constructed,
\begin{equation}
X_{ij} = r_{ij} - {\bar r_j} \ ,
\label{eqn:xij} 
\end{equation}
where ${\bar r_j} = {1 \over N_G} \sum_{i=1}^{N_G} r_{ij}$ is the mean. The
covariance matrix for these quantities is given by
\begin{equation}
C_{jk} = {1 \over N_G} \sum_{i=1}^{N_G} X_{ij} X_{ik} \hskip0.5cm 1 
\leq j \leq N_S \hskip0.5cm 1 \leq k \leq N_S \ .
\end{equation}

It can be shown that the axis (i.e, direction in vector space) 
along which the variance is maximal is
the eigenvector ${\bf e_1}$ of the matrix equation
\begin{equation}
C {\bf e_1} = \lambda_1 {\bf e_1} \ , 
\end{equation}
where $\lambda_1$ is the largest eigenvalue (in fact the variance along
the new axis). The other principal axes and eigenvectors obey similar
equations. It is convenient to sort them in decreasing order (ordering by
variance), and to
quantify the fractional variance by
$\lambda_\alpha/\sum_\alpha\lambda_\alpha$. The matrix of all the
eigenvectors forms a new set of orthogonal axes which are ideally suited
to an efficient description of the data set using a truncated
eigenvector matrix employing only the first $P$ eigenvectors
\begin{equation}
U_P=\{e_{jk}\} \hskip1cm 1 \leq k \leq P \hskip1cm 1 \leq j \leq N_S \ ,
\end{equation}
where $e_{jk}$ is the $j$th component of the $k$th eigenvector.  
The first few eigenvalues account for
most of the variation in the data, and  
the higher eigenvectors contain  mostly the noise (e.g. Folkes, Lahav \& Maddox 1996).
The projection vector
${\bf z}$ onto the $M=N_S$ principal components can be found from (here ${\bf x}$ and
${\bf z}$ are row vectors):
\begin{equation}
{\bf z} = {\bf x} U_M \ .  
\end{equation}
Multiplying by the inverse, the spectrum is given by
\begin{equation}
{\bf x} = {\bf z} {U_M}^{-1} = {\bf z} {U_M}^t \ , 
\end{equation}
since $U_M$ is an orthogonal matrix by definition. However, using only
$P$ principal components, the reconstructed spectrum would be
\begin{equation}
{\bf x}_{rec} = {\bf z} {U_P}^t \ , 
\end{equation}
which is an approximation to the true spectrum.

\begin{figure}
\centering
\includegraphics[width=0.8\textwidth,height=1.0\textwidth]{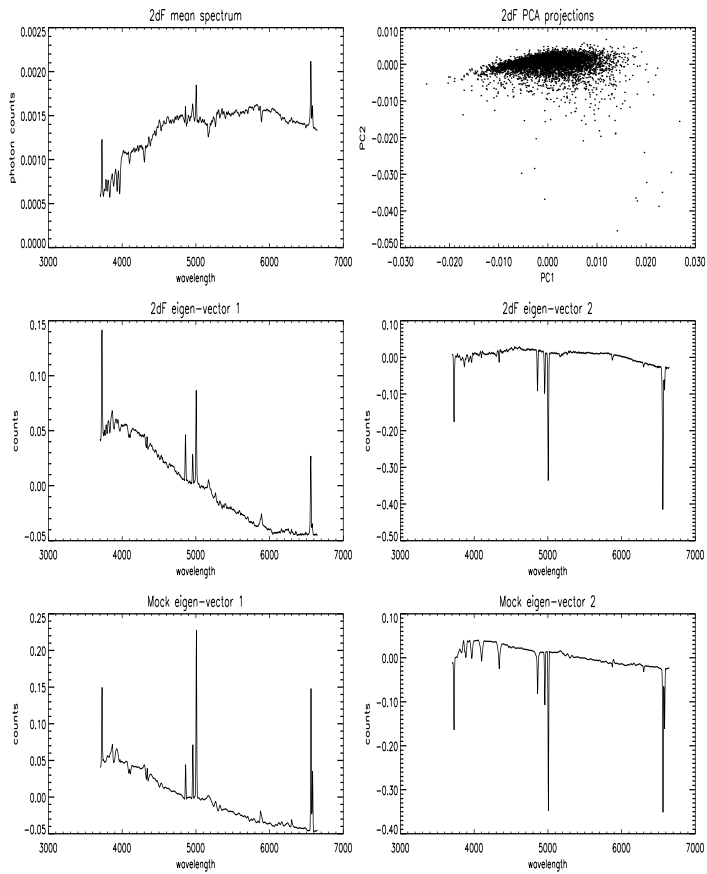}
\caption[]{
The mean spectrum of the 2dF sample 
the first and second eigenvectors
(with associated variances of 50\% and 11 \%, respectively), 
and the projections into the $(pc1,pc2)$ space.
For comparison we show the two eigenvectors of the mock sample. 
\label{pca}}
\end{figure}

The eigenvectors onto which we project the spectra can be viewed as 
`optimal filters' of the spectra, in analogy with other 
spectral diagnostics such as colour filter or spectral index.
Finally, we note that there is some freedom of choice as to whether 
to represent a spectrum as a vector of fluxes or of photon counts. The
decision will affect the resulting principal components, as a representation 
by fluxes will give more weight to the blue end of a spectrum than 
a representation by photon counts. 
Figure~\ref{pca} shows the mean spectrum for the 2dF sample,
the first and second eigenvectors in the 2dF and the mock samples,
and the $(pc1, pc2)$ projections for the 2dF sample.
Given the observational uncertainties described above and the
astrophysical unknowns, the similarity of the real and mock eigenvectors
is quite  remarkable.

\section{The Information Bottleneck (IB)}

Here we summarize  the Information Bottleneck 
(IB) method of Slonim et al. 2000 (SSTL), and  suggest some extensions.
The IB approach is based on 
the method of Tishby, Pereira \& Bialek (1999), 
which
has been successfully applied to the
analysis of neural codes, linguistic data
and 
classification of text documents. In the
latter case, for example, one may see an analogy between an ensemble of
galaxy spectra and a set of text documents. The words in a document
play a similar role to the wavelengths of photons in a galaxy
spectrum, i.e. the frequency of occurrence of a given word in a given
document is equivalent to the number of photon counts at a given
wavelength in a given galaxy spectrum. In both cases, the specific
patterns of these occurrences may be used in order to classify the
galaxies or documents.

\subsection{``Euclidean'' Classification}
\label{alg_simple}

We may gain some intuition into the IB method by first considering 
 standard clustering algorithms. Suppose we
start from Bayes' theorem, where the probability for a class $\c$ for
a given galaxy $\g$ is
\begin{equation}
p(\c|\g)  \propto 
p(\c) \; p(\g|\c)\;,  
\end{equation}
and $p(c)$ is the prior probability for class $c$.  As a simple ad-hoc
example, we can take the conditional probability $p(\g|\c)$ to be a
Gaussian distribution with variance $\sigma^2$
\begin{equation}
p(\g|\c)  = \frac{1}{\sqrt{2 \pi}\sigma}\; 
\exp (-\frac{1}{2 \sigma^2} D_{E}^{2})\;,  
\end{equation}
with the Euclidean distance $D_{E}$:
\begin{equation}
D_{E} = \sqrt{\sum_\l [p(\l|c) -p(\l|g)]^2}\;. 
\end{equation}

The variance $\sigma^2$ may be due
cosmic scatter as well as noise.
Hence $\sigma$ can be viewed as the
`resolution' or the effective `size' of the class in the
high-dimensional representation space.  We note that the Euclidean
distance is commonly used in supervised spectral classification using
`template matching' (e.g. Connolly et al. 1995; Benitez 2000), in
which galaxies are classified by matching the observed spectrum with a
template obtained either from a model or from an observed standard
galaxy. By comparing the IB  method with  this ``Euclidean algorithm'', we
find that the IB  approach yields better class boundaries and preserves
more information for a given number of classes.

\subsection{Mutual Information and the Bottleneck}

In the following we denote the set of galaxies by $\G$ and the array
of wavelength bins by $\Lm$.  We view the ensemble of spectra as a
joint distribution $p(\g,\l)$, which is the joint probability of
observing a photon from galaxy $\g \in \G$ at a wavelength $\l \in
\Lm$.  We normalize the total photon counts in each spectrum (galaxy)
to unity, i.e. we take the prior probability $p(\g)$ of observing a
galaxy $\g$ to be uniform: $p(\g) = \frac{1}{N_{G}}$, where $N_{G}$ is
the number of galaxies in this sample $\G$.  This view of the ensemble
of spectra as a conditional probability distribution function enables
us to undertake the information theory-based approach that we describe
in this section.  Our goal is to group the galaxies into classes that
preserve some objectively defined spectral properties.  Ideally, we
would like to make the number of classes as small as possible (i.e. to
find the `least complex' representation) with minimal loss of the
`important' or `relevant' information.  In order to do this
objectively, we need to define formal measures of `complexity' and
`relevant information'.


The prior probability for a specific class $\c$ is given by
\begin{equation}
\label{r1}
p(\c) =\sum_{\g} p(\g) p(\c|\g) ~.
\end{equation}

We can also write down:
\begin{equation}
\label{r2}
p(\l|\c) = \sum_{\g}p(\l|\g)p(\g|\c) ~,
\end{equation}
where $p(\l|\c)$ can be clearly interpreted as the spectral density
associated with the class $\c$.

The {\em mutual information} between two variables can be shown (see
e.g. Cover \& Thomas 1991) to be given by the amount of uncertainty in
one variable that is removed by the knowledge of the other one, 
for example for the pair $(\G, \Lm)$:
\begin{equation}
\label{MI1}
I(\G;\Lm) =  
\sum_{\g,\l} p(\g,\l) \log \frac {p(\g,\l)}{p(\g)p(\l)} ~ = 
\sum_{\g,\l} p(\g) p(\l|\g) \log \frac {p(\l|\g)}{p(\l)} ~,
\end{equation}
using $p(\g,\l) = p(\g)p(\l|\g)$. 
It is easy to see that $I(\G;\Lm)$ is symmetric and non-negative, and
is equal to zero if and only if $\g$ and $\l$ are independent.  
Similarly we can define
the mutual information between a set of {\em galaxy classes} $\C$
and the {\em spectral wavelengths} $\Lm$,
$I(\C;\Lm) = \sum_{\c,\l} p(\c) p(\l|\c) \log \frac {p(\l|\c)}{p(\l)}$,
and 
the mutual information between the {\em classes and 
the galaxies}
$I(\C;\G) = \sum_{\c,\g} p(\g) p(\c|\g) \log \frac {p(\c|\g)}{p(\c)}$.

A basic theorem in information theory, known as {\em data processing
inequality}, states that no manipulation of the data can increase the
amount of (mutual) information given in that data.  Specifically this
means that by grouping the galaxies into classes one can only {\em
lose} information about the spectra, i.e.  $I(\C;\Lm) \leq I(\G;\Lm)$.

The problem can be formulated as follows: how
do we find classes of galaxies that maximize $I(\C;\Lm)$, under a
constraint on $I(\C;\G)$?  
In effect we pass the information that
$\G$ provides about $\Lm$ through a ``bottleneck'' formed by the
classes in $\C$.
The classes $\C$ are forced to extract the relevant
information between $\G$ and $\Lm$.
Hence the name 
{\it information bottleneck method} 
\footnote{the `bottleneck' is analogous to the `hidden layer'
between the input and output layers in Artificial Neural Networks 
(see e.g. Lahav et al. 1996).}. 

Under this formulation, the {\em optimal} classification is given by
maximizing the functional
\begin{equation}
\label{varprin}
\L [p(\c|\g)] =  I(\C ; \Lm) - \beta^{-1} I(\C; \G) ,
\end{equation}
where $\bb$ is the Lagrange multiplier attached to the complexity
constraint.  For $\beta \rightarrow 0$ our classification is as
non-informative (and compact) as possible --- all galaxies are
assigned to a single class.  On the other hand, as $\beta \rightarrow
\infty$ the representation becomes arbitrarily detailed.  By varying
the single parameter $\beta$, one can explore the tradeoff between the
preserved meaningful information, $I(\C;\Lm)$, and the compression
level, $I(\C;G)$, at various resolutions.

The optimal assignment
that maximizes Eq. (\ref{varprin}) satisfies the equation
\begin{equation}
\label{r3}
p(\c|\g)
= {{p(\c)}\over Z(\g,\beta)}
\exp(- \beta D_{KL}) ~,
\end{equation}
where $Z(g,\beta)$ is the common normalisation (partition) function
\footnote{
We note that $\beta$ here is analogous to the inverse temperature in
the Boltzmann's distribution function.}.  The value in the exponent can
be considered the relevant ``distortion function'' between the class
and the galaxy spectrum. It turns out to be the familiar cross-entropy
(also known as the `Kullback-Leibler divergence', e.g. Cover \& Thomas
1991), defined by
\begin{equation}
\label{KL}
D_{KL}\left[p(\l|\g) \Vert p(\l|\c)\right] = \sum_{\l} p(\l|\g) \log
{{p(\l|\g)}\over{p(\l|\c)}} ~.
\end{equation}
Note that Eqs. (\ref{r1}, \ref{r2}, \ref{r3})
must be solved together in a {\em self-consistent} manner.
We can also see now  the analogy 
with the `Euclidean equations' (section 4.1), i.e. 
between $D_{KL}$ and $D_E$,
and between $\beta$ and $\sigma^2$.
The IB approach is obviously far more `principled'.

 \subsection{The Agglomerative IB Algorithm} 
\label{alg}

In practice
we actually used a special case of the  algorithm,
based on a bottom-up {\em merging} process.
This algorithm generates ``hard'' classifications, i.e.  every galaxy
$\g \in \G$ is assigned to exactly one class $\c \in \C$. Therefore,
the membership probabilities $p(\c|\g)$ may only have values of $0$ or
$1$.  Thus, a specific class $\c$ is defined by the following
equations, which are actually the ``hard'' limit $\beta
\rightarrow \infty$ of the general self-consistent Eqs. (\ref{r1},
\ref{r2}, \ref{r3}) presented previously,
\beq{hard1}
\left\{
\begin{array}{l}
p(\c)  = \sum_{\g \in \c}{p(\g)} \\\\
p(\l|\c)  = \frac{1}{p(\c)} \sum_{\g \in \c} p(\l|\g)p(\g) \\\\
p(\c|\g) = \left\{ \begin{array}{ll} 1 & \mbox{if $\g \in \c$} \\ 0 &
      \mbox{otherwise} \end{array} \right. \\
\end{array}
\right.
\eeq
where for the second equation we used Bayes' theorem, $p(\g|\c) =
\frac{1}{p(\c)}p(\c|\g)p(\g)$.

The algorithm starts with the trivial solution, where $\C \equiv \G$
and every galaxy is in a class of its own. In every step two classes
are merged such that the mutual information $I(\C;\Lm)$ is maximally
preserved.  
Note that this algorithm
naturally finds a classification for any desired number of classes
with no need to take into account the theoretical constraint via
$\beta$ (Eq. \ref{varprin}).  This is due to the fact that the
agglomerative procedure contains an inherent {\em algorithmic}
compression constraint, i.e. the merging process 
(for more details see SSTL).

\subsection{ IB Classification Results}

\begin{figure}
\centering 
\includegraphics[width=0.8\textwidth,height=0.4\textwidth ]{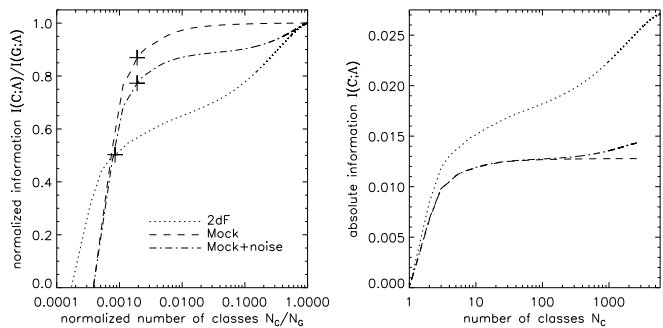}
\caption[]
{Left Panel: The fractional information measure
($I(\C;\Lm)/I(\G;\Lm)$) vs. the fractional number of classes
($N_{C}/N_{G}$). The crosses mark the points corresponding to five
classes, used in the remainder of this paper.  Right Panel: The
`absolute' information content $I(\C;\Lm)$ as a function of the number
of classes $N_C$. (From SSTL 2000.)  }
\label{fig:info}
\end{figure}

We now apply the IB algorithm to both the 2dF and the mock data.
Recall that our algorithm begins with one class per galaxy, and groups
galaxies so as to minimize the loss of information at each
stage. Figure~\ref{fig:info} shows how the information content of the
ensemble of galaxy spectra decreases as the galaxies are grouped
together and the number of classes decreases. In the left panel, we
show the `normalized' information content $I(\C;\Lm)/I(\G;\Lm)$ as a
function of the reduced complexity $N_{C}/N_{G}$, where $N_{G}$ is the
number of galaxies in the ensemble and $N_{C}$ is the number of
classes.  
Remarkably, we find that if we keep about five classes, about
$85$ and $75$ percent of the information is preserved for the mock and
mock+noise simulations, respectively. This indicates that the
wavelength bins in the model galaxy spectra are highly correlated.
In contrast with the mock samples, for the 2dF catalogue, only about
$50$ percent of the information is preserved by five classes 
\footnote {We note that 
galaxy images can be reliably classified 
by morphology 
into no more than 7 or so classes
(e.g. Lahav et al. 1995; Naim et al. 1995a). }.
This discrepancy may be partially due to the influence on the real spectra
of more complicated physics than what is included in our simple
models. It could also be due to systematic observational errors 
mentioned earlier.

\begin{figure}
\centering
\includegraphics[width=0.9\textwidth,height=1.1\textwidth]{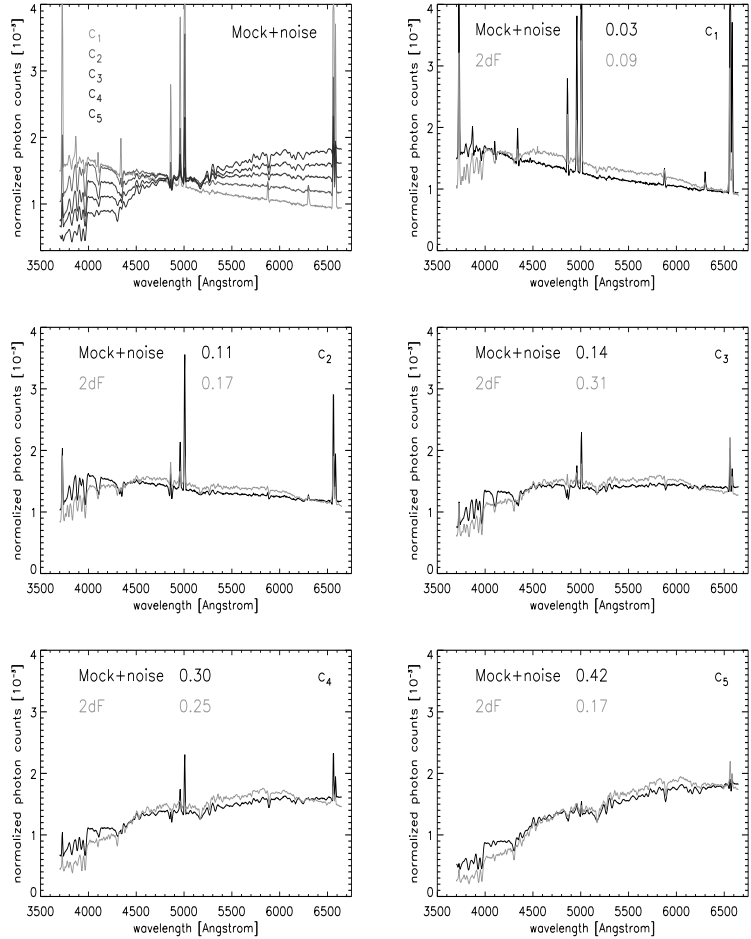}
\caption[]
{Mean spectra of the five IB classes for the 2dF and mock+noise
catalogues. The fraction $p(c_j)$ of galaxies that are members of each
class $c_j$ is indicated. The matching between the classes obtained
for 2dF and the mock catalogue was determined by minimizing the
average  `distance' between the pairs. (From SSTL 2000.)
\label{fig:spectra5class}}
\end{figure}

We now present the results obtained for
five classes.  
Figure~\ref{fig:spectra5class} shows the representative
spectra for these five classes for both the 2dF and mock+noise
catalogues. The corresponding five spectra for the noise-free mock
data were very similar to the mock+noise spectra shown. We `matched'
each of the classes obtained for the 2dF data with one from the
mock+noise data by minimizing the average `distance' between
the pairs. The classes are then ordered by their mean $B-V$
colour. Note that the five classes produced by the algorithm appear
similar for both catalogues --- there was certainly no guarantee that
this would be the case.

It is also interesting to examine the relative fractions of galaxies
in each class, $p(c)$, for the observed and mock catalogues. These
values are given on the appropriate panels of
Figure~\ref{fig:spectra5class}. 
More generally, we can see that the algorithm is sensitive to the
overall slope (or colour) of the spectrum, and also to the strength of
the emission lines. The classes clearly preserve the familiar physical
correlation of colour and emission line strength; the five classes
form a sequence from $\c_1$, which has a blue continuum with strong
emission lines, to $\c_5$, with red continuum with no emission lines.
It is interesting to compare the mean spectrum
of $\c_1$ with the spectrum of the Sm/Irr galaxy NGC449, and $\c_5$
with the Sa galaxy NGC775 from Figure 2a of Kennicutt (1992).
Apparently, the $\c_1$ class corresponds to late type galaxies
(Sm/Irr) and $\c_5$ to early types (Sa-E).
In order to gain a better understanding of the IB classes, we also use
the noise-free mock catalogue and investigate the physical properties
of the galaxies in each class as given by the same models that we use
to produce the spectra. 
The strongest trend is of $B-V$ colour and
present-to-past-averaged star formation rate 
(see Figure 5 in SSTL).


\subsection{Comparison of IB with PCA} 


It is  interesting to see where
the IB classes reside in the space of the PCA projections. 
The  5 IB classes form fairly
well-separated ``clumps'' in PC1-PC2 space, and that to a first
approximation, the IB classification is along PC1
(see Figure 14 in SSTL).
The
PCA-space of the IB clumps looks quite different from the partitioning
(based on training sets) given in Folkes et al. (1999),
which was mainly based on emission and absorption lines
(rather than on the continuum).  
It has been
shown (Ronen et al. 1999) that PC1 and PC2 are correlated with colour
and emission line strength, and the sequence from $\c_1$-$\c_5$ is
again sensible in this context.

\subsection{Extensions of the IB Approach} 

\subsubsection{Unsupervised `Wavelength Grouping'}

We can ask a different question using the IB tool:
What set $\C_{\Lm} $ of wavelength combinations
(`filters') are the best indicators of the galaxy identity ?

This can be done by simply interchanging $\g$ and $\l$ in eq. ~\ref{hard1}:

\beq{wave1}
\left\{
\begin{array}{l}
p(\c_{\Lm})  = \sum_{\l \in \c_{\Lm}}{p(\l)} \\\\
p(\g|\c_{\Lm})  = \frac{1}{p(\c_{\Lm})} \sum_{\l \in \c_{\Lm}} p(\g|\l)p(\l) \\\\
p(\c_{\Lm}|\l) = \left\{ \begin{array}{ll} 1 & 
      \mbox{if $\l \in \c_{\Lm}$} \\ 0 &
      \mbox{otherwise} \end{array} \right. \\
\end{array}
\right.
\eeq

One difference is that while we have taken $p(\g) = 1/N_G$,
here $p(\l)$ is the mean spectrum of the sample. 
Our preliminary results indicate that 10 combinations of spectral lines
retain 68 \% of the information in the case of the 2dF sample,
and 96 \% for the mock sample (Slonim et al., in preparation).

\subsubsection{Supervised `Wavelength Grouping'}

Another approach is to group wavelengths which are the most 
informative about physical parameters, e.g. 
age, star formation rate, etc.
Commonly this question is answered by a more 
intuitive way, e.g. by extracting the H$\alpha$ line
as an indicator of star formation rate.

This is in fact a formal principled solution to the fundamental 
question on the best spectral diagnostics of physical 
parameters. 
For a physical parameter of interest $a$
the relevant set of equations is now:

\beq{wave2}
\left\{
\begin{array}{l}
p(\c_{\Lm})  = \sum_{\l \in \c_{\Lm}}{p(\l)} \\\\
p(a|\c_{\Lm})  = 
\frac{1}{p(\c_{\Lm})} \sum_{\l \in \c_{\Lm}} p(a|\l)p(\l) \\\\
p(\c_{\Lm}|\l) = \left\{ \begin{array}{ll} 1 & 
      \mbox{if $\l \in \c_{\Lm}$} \\ 0 &
      \mbox{otherwise} \end{array} \right. \\
\end{array}
\right.
\eeq

We shall show elsewhere the implementation of these
`wavelength grouping' algorithms. 
The next section is another approach for relating spectral 
features to physical parameters.

\section{Maximum Likelihood and the Fisher Matrix (FM) Approach} 

In the case that an underlying astrophysical model 
for the spectrum is assumed, one may find the best fit parameters,
by Maximum Likelihood.
Heavens, Jimenez \& Lahav (HJL 2000)  
presented a Maximum Likelihood method 
with radical linear compression of the datasets.
In the case 
that the  noise in the data is independent of the parameters, one
can form $M$ linear combinations of the data which contain as much
information about all the parameters as the entire dataset, in the
sense that the Fisher information matrices are identical; i.e. the
method is lossless.  
When
the noise is dependent on the parameters (as in the case of galaxy spectra),
the method
is not precisely lossless, but the errors 
increase by a very modest factor.  
This data compression offers the possibility of a large increase in
the speed of determining physical parameters.  This is an important
consideration as datasets of galaxy spectra reach $\sim 10^6$ in size, and
the complexity of model spectra increases.  In addition to this
practical advantage, the compressed data may offer a classification
scheme for galaxy spectra which is based rather directly on physical
processes.

\subsection{The FM Compression Method}

Here we represent a spectrum as a vector $\bx_i$, $i=1,\ldots,N_S$ (e.g. a set
of fluxes at different wavelengths).  These measurements include a
signal part, which we denote by $\bbmu$, and noise, $\bn$:
\begin{equation}
\bx = \bbmu + \bn
\end{equation}
Assuming the noise has zero mean, $\langle \bx\rangle = \bbmu$.  The
signal will depend on a set of physical parameters $\{\theta_\alpha\}$, which
we wish to determine.  For galaxy spectra, the parameters may be, for
example, age, magnitude of source, metallicity and some parameters
describing the star formation history.  Thus, $\bbmu$ is a noise-free
spectrum of a galaxy with certain age, metallicity etc.

The noise properties are described by the noise covariance matrix,
$\bC$, with components $C_{ij} = \langle n_i n_j\rangle$.  If the
noise is Gaussian, the statistical properties of the data are
determined entirely by $\bbmu$ and $\bC$.  In principle, the noise can
also depend on the parameters.  For example, in galaxy spectra, one
component of the noise will come from photon counting statistics, and
the contribution of this to the noise will depend on the mean number
of photons expected from the source.

The aim is to derive the parameters from the data. If we assume
uniform priors for the parameters, then the a posteriori probability
for the parameters is the likelihood,  which for Gaussian noise is
\ba
{\cal L}(\theta_\alpha) &=& {1\over (2\pi)^{N/2}
\sqrt{\det(\bC)}}
\times \nn & & 
\exp \left[-{1\over 2}\sum_{i,j} (x_i -
\mu_i)\bC^{-1}_{ij} (x_j-\mu_j)\right].
\ea
One approach is simply to find the (highest) peak in the likelihood,
by exploring all parameter space, and using all $N_S$ pixels.  The
position of the peak gives estimates of the parameters which are
asymptotically (low noise) the best unbiased estimators.
This is therefore the best we can do.  The maximum-likelihood
procedure can, however, be time-consuming if $N_S$ is large, and the
parameter space is large.  
The aim is 
to reduce the $N_S$ numbers to a smaller number, without increasing the
uncertainties on the derived parameters $\theta_\alpha$.  To be
specific, we try to find a number $M<N_S$ of linear combinations of the
spectral data \bx\ which encompass as much as possible of the
information about the physical parameters.  We find that this can be
done losslessly in some circumstances; the spectra can be reduced to a
handful of numbers without loss of information.  The speed-up in
parameter estimation is about a factor $\sim 100$.

In general, reducing the dataset in this way will lead to larger error
bars in the parameters.  To assess how well the compression is doing,
consider the behaviour of the (logarithm of the) likelihood function
near the peak.  Performing a Taylor expansion and truncating at the
second-order terms,
\begin{equation}
\ln{\cal L} = \ln{\cal L}_{\rm peak} +{1\over 2}  {\partial^2 \ln {\cal L}\over
\partial \theta_\alpha \partial
\theta_\beta}\Delta\theta_\alpha\Delta\theta_\beta.
\end{equation}
Truncating here assumes that the likelihood surface itself is
adequately approximated by a Gaussian everywhere, not just at the
maximum-likelihood point.   The actual likelihood surface will vary
when different data are used;  on average, though, the width is set by 
the (inverse of the) Fisher information matrix:
\begin{equation}
\bF_{\alpha\beta} \equiv - \left\langle {\partial^2 \ln {\cal L}\over
\partial \theta_\alpha \partial \theta_\beta} \right\rangle
\end{equation}
where the average is over an ensemble with the same parameters but different
noise.  
For more discussion on the Fisher matrix see
Tegmark, Taylor \& Heavens (1997).  

In practice, some of the data may tell us very little about the
parameters, either through being very noisy, or through having no
sensitivity to the parameters.  So in principle we may be able to
throw some data away without losing very much information about the
parameters.  Rather than throwing individual data away, we can do
better by forming linear combinations of the data, and then throwing
away the combinations which tell us least.  To proceed, we first
consider a single linear combination of the data:
\begin{equation}
y_1 \equiv \bbf_{1}^t \bx
\end{equation}
for some weighting vector $\bbf_{1}$ ($t$ indicates transpose).  
The idea is
to find a weighting which captures as much information about a
particular parameter, $\theta_1$.  
It turns out that the solution (properly normalised) is:
\begin{equation}
\bbf_{1} = {\bC^{-1} \bbmu_{,1}\over \sqrt{\bbmu_{,1}^t
\bC^{-1}\bbmu_{,1}}},
\label{Evector1}
\end{equation}
where $\bbmu_{,1} = {\partial {\bbmu} \over {\partial \theta_1} } $.
Our compressed datum is then a  single number 
$y_{1}=\bbf_{1}^{t} \bx$.  
Normally one has several parameters to estimate
simultaneously, and this introduces substantial complications into the
analysis.  How can we generalise the single-parameter estimate above
to the case of many parameters ?  
We proceed by finding a second number
$y_2 \equiv \bbf_2^t \bx$, uncorrelated with $y_1$ by construction.
It is also required that 
$y_2$ captures as much information as possible about the
second parameter $\theta_{2}$.
The vectors $\bbf_1$, $\bbf_2$, etc. are analogous to the eigenvectors
in the PCA approach, and can also be viewed as `optimal filters'
of the spectra.

Since, by construction, the numbers $y_m$ are uncorrelated, the
likelihood of the parameters is obtained by multiplication of the
likelihoods obtained from each statistic $y_m$.  The $y_m$ have mean
$\langle y_m \rangle = \bbf_m^t \bbmu$ and unit variance, so the
likelihood from the compressed data is simply 
\be
\ln{\cal L}(\theta_\alpha) = {\rm constant} - \sum_{m=1}^{M}
{(y_m-\langle y_m\rangle)^2\over 2}.
\ee

In practice, one does not know beforehand what the true solution is,
so one has to make an initial guess (`a fiducial model') 
for the parameters.  
One can iterate: choose a fiducial model; use
it to estimate the parameters, and then repeat, using the 
estimated parameters as the fiducial model.

\subsection{Example - Estimating Galaxy Age}

\begin{figure}
\centering
\includegraphics[width=0.5\textwidth,angle=270]{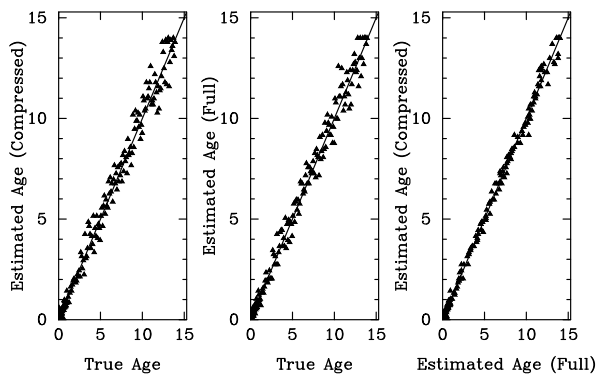}
\caption[]{The effect of the fiducial model on recovery
of the parameters.  Here a single fiducial model is chosen (with age 9 
Gyr), and ages recovered from many true galaxy spectra with ages
between zero and 14 Gyr. The left panel shows the recovered age from the
two numbers $y_1$ and $y_2$ (with age and normalisation weightings),
plotted against the true model age.  The middle panel shows how well the 
full dataset (with S/N $\ls 2$) can recover the parameters.  The right
panel shows the estimated age from the $y_1$ and $y_2$ plotted against 
the age recovered from the full dataset, showing that the compression
adds very little to the error, even if the fiducial model is very
wrong.  Note also that the scatter increases with age;  old galaxies
are more difficult to date accurately. (from HJL 2000.)}
\label{SN2F}
\end{figure}

An example result from the
two-parameter problem is shown in Fig. \ref{SN2F}.  Here the ages and
normalisations (of the star-formation-rate) 
of a set of model galaxies with S/N $\ls 2$ are
estimated, using a common (9 Gyr) galaxy as the fiducial model. We see
that the method is successful at recovering the age, even if the
fiducial model is very badly wrong.  There are errors, of course, but
the important aspect is that the compressed data 
do almost as well as the full data set.

\subsection{Comparison of the FM with PCA}

HJL contrasted the Fisher Matrix method with PCA, by comparing the
eigenvectors of the two methods.  PCA is not lossless unless all
principal components are used, and compares unfavourably in this
respect for parameter estimation.  However, one requires a theoretical
model for the Fisher method; PCA does not require one, needing instead
a representative ensemble for effective use.  Other, more ad hoc,
schemes consider particular features in the spectrum, such as
broad-band colours, or equivalent widths of lines (e.g. Worthey 1994).
Each of these is a ratio of linear projections, with weightings given
by the filter response or sharp filters concentrated at the line.
There may well be merit in the way the weightings are constructed, but
they will not in general do as well as the optimum weightings
presented here.

\section{Discussion}
\label{sec:conclusions}

We summarized three recently used methods for compression and
classification of galaxy spectra.  Studies of the PCA method for
galaxy spectra have shown that only 3-8 Principal Components are
required to represent 2dF-like spectra.  
The PCA is indeed very
effective for data compression, but if one wishes to break the
ensemble into classes it requires a further step based on a training
set (e.g. Bromley et al. 1998, Folkes et al. 1999).  An alternative
approach to dividing the PCA space into classes is to combine the
projected PCs into a one-parameter (sequence-like) model which
represents meaningful features of the spectra, while minimizing
instrumental effects (Madgwick, Lahav \& Taylor 2000, in this volume).
In a way, this is related to the old and deeper question, whether the galaxy
population forms a sequence or is made of  distinct classes.
PCA can be generalized to more powerful linear projections,
e.g. projection pursuit (Friedman and Tukey 1974) or
to nonlinear projections that maximize statistical independence, such as
Independent Component Analysis (ICA; Bell and Sejnowski 1995).
These methods provide a low dimensional representation, or
compression, in which one might hope to
identify the relevant structure more easily.

Unlike PCA,
the Information Bottleneck (IB) method of SSTL
is non-linear, and it naturally yields a principled partitioning
of the galaxies into classes.  These classes are obtained such that
they maximally preserve the original information between the galaxies
and their spectra.
The IB method  makes no
model-dependent assumptions on the data origin, nor about the
similarity or metric among data points.
The analysis of 2dF and mock spectra suggests that  
5-7 spectral classes preserve most of the information.

If, on the other hand, 
one has a well defined physical model for galaxy spectra, then 
it is appropriate to estimate parameters of interest 
(e.g. age and star-formation rate) by Maximum Likelihood.
This can be done directly using all the spectral bins, or via a linear 
compressed version of the data designed to preserve information 
in the sense of `Fisher Matrix' (FM) with respect
to  the physical parameters of interest, as shown by HJL.
We emphasize that both the PCA and FM methods are linear, 
while the IB is non-linear.
Unlike the FM method,
PCA and the IB methods are model-independent, and they require ensembles
of spectra. 
The IB `supervised wavelength grouping' (section 4.1) is close conceptually 
to the approach of the FM method.

Although not discussed here, another non-linear approach of
identifying classes of objects in a parameter-space (based on a
training set) is by utilising Artificial Neural Networks (e.g. used
for morphological classification of galaxies; Naim et al. 1995b; Lahav
et al. 1996).

The above methods illustrate that 
automatic classification of millions of galaxies is feasible.
As `the proof of the pudding is in the eating',
these methods should be judged eventually by their `predictive power'.
In particular if the spectral diagnostics can reveal new astrophysical
features and remove e.g. the age-metallicity degeneracy.
Another important application
is related to the global distribution of galaxies,
i.e. luminosity functions per spectral type 
(e.g. Bromely et al. 1998, Folkes et al. 1999) 
and large scale clustering per spectral class (or physical parameter),
with the obvious implications for galaxy formation and theories of 
biasing. Future work may include:
\begin{itemize}
\item
On the data side - improving flux calibration (to have a reliable continuum),
and quantifying selection effects such as 
fibre aperture bias.
\item
On the modelling side - improving models for emission lines, dust etc. 
\item 
On the algorithmic side - exploring
new unsupervised and supervised methods.
\end{itemize}

\section{Acknowledgments}
I thank A. Heavens, R. Jimenez, 
D. Madgwick, 
N. Slonim, R. Somerville, N. Tishby, 
and the 2dFGRS team for their contribution to the work presented here.
I also thank A. Banday and S. Zaroubi for suggesting to me to review this topic.



\clearpage
\addcontentsline{toc}{section}{Index}
\flushbottom
\printindex


\begin{thebibliography}{7}
%
\addcontentsline{toc}{section}{References}

\bibitem{b1}  Bailer-Jones C.A.L., Irwin M., Gilmore G., von Hippel T., 1997, MNRAS, 292, 157 
\bibitem{bs} Bell A.J., Sejnowski T.J., 1995, Neural Computation, 7, 1129
\bibitem{bn} Benitez N., 2000, ApJ, 536, 571
\bibitem{Bromley98}Bromley B., Press W., Lin H.,
  Kirschner R., 1998, ApJ, 505, 25
\bibitem{b5} Bruzual G., Charlot S., 1993, ApJ, 405, 538
\bibitem{b4} Bruzual G., Charlot S., 1996, Galaxy Isochrone Synthesis Spectral
 Evolution Library, Multi Metallicity Version (GISSEL96)
\bibitem{b6} Colless M.M., 1998, in Morganti R., Couch W.J., eds, ESO/Australia workshop, Looking Deep in the Southern Sky, Springer Verlag, Berlin  p. 9 
\bibitem{Connolly95}Connolly A., Szalay A.,
  Bershady M., Kinney A., Calzetti D., 1995,  AJ, 110,
  1071
\bibitem{ct} Cover, T.M., Thomas, J.A. 1991, Elements of Information Theory, John Wiley \& Sons, New York
\bibitem{b8} Fioc M., Rocca-Volmerange B., 1997, A\&A, 326, 950
\bibitem{b9} Folkes S.R., Lahav O., Maddox S.J., 1996, MNRAS, 283, 651
\bibitem{Folkes99}Folkes S., Ronen S., Price I.,
  Lahav O., Colless M., Maddox S.~J., Deeley K.~E., Glazebrook K.,
  Bland-Hawthorn J., Cannon R.~D., Cole S., Collins C.~A., Couch W., Driver
  S.~P., Dalton G., Efstathiou G., Ellis R.~S., Frenk C.~S., Kaiser N., Lewis
  I.~J., Lumsden S.~L., Peacock J.~A., Peterson B.~A., Sutherland W., Taylor
  K., 1999,  MNRAS, 308, 459
\bibitem{b11} Francis P.J., Hewett P.C., Foltz G.B., Chaffee, F.H., 1992, ApJ, 398, 476
\bibitem{ft} Friedman J.H., Tukey J.W., 1974, IEEE Trans. Comput. C(23), 881
\bibitem{Galaz98}Galaz G., de~Lapparent V.,
  1998, A\&A,  332, 459
\bibitem{Glazebrook98}Glazebrook K., Offer
  A., Deeley K., 1998,  ApJ, 492, 98
\bibitem{grv} Guiderdoni B., Rocca-Volmerange, B., 1987, A\&A, 186, 1
\bibitem{hjl} Heavens A., Jimenez R., Lahav, O., 2000, MNRAS,
317, 965 (HJL)
\bibitem{} Humason, M.L., 1936, ApJ, 83, 18 
\bibitem{k92} Kennicutt R.C. 1992, ApJS, 388, 310
\bibitem{} Kochanek, C.S., Pahre, M.A., Falco, E.E., 2000, 
astro-ph/0011458
\bibitem{} Lahav, O. et al., 1995, Science, 267, 859
\bibitem{} Lahav, O., Naim, A., Sodre, L., Storrie-Lombardi, M.C., 1996, 
MNRAS, 283, 207
\bibitem {MLT} Madgwick, D.S., Lahav, O., Taylor, K. (and the 2dFGRS team), 
2000, in proceedings of the MPA/ESO workshop Mining the Sky,
eds. A. Banday et al., Springer-Verlag, this volume  
\bibitem{} Morgan, W.W, Mayall, N.U., 1957, PASP, 69, 291 
\bibitem{Murtagh87}{Murtagh} F., {Heck} A.,
  1987.\newblock {\it Multivariate data analysis}, Astrophysics and Space
  Science Library, Reidel, Dordrecht.
\bibitem{}Naim A.,  et al., 1995a, MNRAS, 274, 1107
\bibitem{}Naim A., Lahav, O., Sodre, L., Storrie-Lombardi, M.C., 
 1995b, MNRAS, 275, 567
\bibitem{Ronen99}Ronen R.~T.,
  Aragon-Salamanca A., Lahav O., 1999,  MNRAS,  303, 284
\bibitem{ST00} Slonim N., Somerville, R. Tishby N., Lahav, O.
2000, MNRAS, in press, astro-ph/0005306 (SSTL) 
\bibitem{b24} Sodr\'e L. Jr., Cuevas H., 1997, MNRAS, 287, 137
\bibitem{thesis} Somerville R.S., 1997, PhD Thesis, Univ. California, Santa Cruz
\bibitem{sp} Somerville R.S., Primack, J.R., 1999, MNRAS, 310, 1087
\bibitem{TTH97}Tegmark M., Taylor A., Heavens
  A., 1997, ApJ, 480, 22
\bibitem{Worthey94}Worthey G., 1994, ApJSS, 
  95, 107
\bibitem{TPB99} Tishby N., Pereira F.C., Bialek W., 1999, Proc. of the 37th Allerton Conference 
on Communication and Computation
\bibitem{}
van den Bergh, S.,{\it Galaxy Morphology and Classification}, 1998, 
Cambridge University Press, Cambridge

\end{thebibliography}
\end{document}